\documentstyle[12pt,epsf]{article}

\def\beq{\begin{equation}}   \def\eeq{\end{equation}}

\newcommand{\gsim}{\lower.7ex\hbox{$\;\stackrel{\textstyle>}{\sim}\;$}}
\newcommand{\lsim}{\lower.7ex\hbox{$\;\stackrel{\textstyle<}{\sim}\;$}}

\begin{document}

\def\lsim{\mathrel{\rlap{\lower3pt\hbox{\hskip0pt$\sim$}}
    \raise1pt\hbox{$<$}}}         %less than or approx. symbol
\def\gsim{\mathrel{\rlap{\lower4pt\hbox{\hskip1pt$\sim$}}
    \raise1pt\hbox{$>$}}}         %greater than or approx. symbol

\begin{titlepage}
\renewcommand{\thefootnote}{\fnsymbol{footnote}}

\begin{flushright}

TPI-MINN-97/22-T\\
UMN-TH-1603/97\\
hep-th/9708060\\

\end{flushright}

\vspace{0.3cm}

\begin{center}
\baselineskip25pt

{\Large\bf 
 Degeneracy  and Continuous
Deformations  of Supersymmetric Domain Walls}

\end{center}

\vspace{0.3cm}

\begin{center}
\baselineskip12pt

\def\thefootnote{\fnsymbol{footnote}}

\vspace{0.3cm}
{\large  M.~Shifman} 

\vspace{0.5cm}

  Theoretical Physics Institute, University of Minnesota, Minneapolis,
MN 54555 USA$^\dagger$ \\[0.5cm]

\vspace{0.1cm}

{\em and}

\vspace{0.3cm}

{Institut f\"{u}r Theoretische Physik III, Universit\"{a}t
Erlangen-N\"{u}rnberg, D-91058 Erlangen, Germany}

\vspace{1.5cm}

{\large\bf Abstract}

 \vspace*{.25cm}
\end{center}

In a wide class of supersymmetric theories
degenerate families of the BPS-saturated
domain walls exist. The internal structure of these walls
can continuously vary, without changing the wall tension.
This is described by hidden parameters
(collective coordinates).
Differentiating with respect to the collective
coordinates one gets a set of the bosonic zero modes
localized on the wall. Neither of them is
related to the spontaneous breaking of any symmetry.
Through the residual 1/2 of supersymmetry each bosonic zero
mode generates a fermionic partner.

\vspace{2.5cm}

\hfill

\begin{flushleft}

July  1997

\vspace{0.25cm}

\rule{2.4in}{.25mm} \\
$^\dagger$ Permanent address.

\end{flushleft}

\end{titlepage}

\section{Introduction}

The central extensions of $N=1$ superalgebras
in four dimensions 
discovered recently \cite{DS1,DS2,KSS,CS}
lead to the existence of the BPS-saturated domain walls
in supersymmetric theories, with rather peculiar properties.
In Ref. \cite{CS} it was noted, in particular,
that one and the same model can have a variety of distinct domain walls
interpolating between the given pair of vacua. 
Here this remark is elaborated.
It will be shown that in a wide class of supersymmetric models, typically,
a (continuously degenerate) family of the domain walls
exist. Each domain wall from the family is labelled by one or more hidden 
parameter(s).
Although the internal structure of each domain wall is different,
they all have one and the same energy density ${\cal E}$.
One can view  the domain walls from this family as
bound states of two (or more) ``basic''
domain walls, with the vanishing binding energy.
In other words, the ``basic''
domain walls do not interact.
In the two-dimensional reductions of the four-dimensional theories
under consideration the domain walls become kinks (solitons).
Our result translates then into a statement
that the basic ``solitons'' do not interact with each other.

The hidden parameters are the collective coordinates of the domain 
wall solutions. The existence of some collective coordinates
is a trivial consequence of the fact that the domain walls spontaneously break
a part of the four-dimensional symmetries: translational invariance in
the $z$ direction and 1/2 of supersymmetry (if the domain walls at hand are
BPS-saturated). Therefore, the ocurrence of a coordinate $z_0$
usually referred to as the wall center, is not surprising.
We will show that similar coordinates survive for each individual
``component'' of the ``composite'' BPS wall. An analogous  situation
takes place  in the two-instanton solution of the Yang-Mills theory.
Each instanton is characterized by its individual center,
so that we have eight collective coordinates associated with
translations, although only four translational symmetries 
of the theory are spontaneously broken
on the solution. The symmetry of the solution
is higher than that of the theory itself.

One can introduce the overall wall center, $Z_0$,
and {\em extra} collective coordinates $R_i$, which have the 
physical  meaning
of, roughly,  relative ``distances'' between different ``components'' of the
wall. If all parameters $R_i$ tend to infinity, the ``basic''  components 
of the wall are
 infinitely separated. The existence of such a limiting solution is trivial.
The solution persists, however, at finite values of $R_i$,
with the same tension ${\cal E}$, independent of $R_i$.

Differentiating with respect to the collective coordinates $R_i$
one generates  zero modes, localized on the wall and associated with
a change in the internal structure of the wall.
 These zero modes are unrelated to the trivial zero mode corresponding   
to the shift of a wall, as a whole, in the $z$ direction.
Since 1/2 of supersymmetry is preserved, each extra bosonic zero
mode will be accompanied by a fermionic counterpartner.

The continuously  degenerate  domain walls 
occur in the models in which the parameters in the superpotential
${\cal W}(\Phi , X, ...)$
are real (or can be made real by an appropriate
transformation  of the  fields $\Phi , X,...$), 
and all extrema of the superpotential
(classical minima of the potential) occur at real values
of the fields $\phi ,\chi$ and so on. (Here $\phi, \chi $, ...
denote the lowest components of the superfields $\Phi , X$, ...).
The class of theories admitting the degenerate families
of the  domain wall solutions
is actually much wider, especially if one includes into consideration
 non-renormalizable and/or non-polynomial superpotentials.
The latter naturally appear in effective low-energy theories,
see e.g. \cite{VY,KS}. 
Since  generalization is quite straightforward,
and will become completely clear from what follows,
I will limit myself here
to the generalized Wess-Zumino (WZ) models \cite{WZ} with
 renormalizable superpotentials, assuming  real values of 
parameters, the more so that  many practically
important problems belong to this class.
 
Assume for definiteness that 
the superpotential ${\cal W}$ has three extrema
(I will call them generically ${\cal M}_1$, ${\cal M}_2$ and
 ${\cal M}_3$ where 
${\cal M}$ stands for a complete set of superfields in the problem at hand.)
These extrema are ordered in such a way that 
${\cal W}({\cal M}_1) <{\cal W}({\cal M}_2) <{\cal W}({\cal M}_3)$.
The energy density of the BPS domain walls is proportional
to  the central
charge in the corresponding transition \cite{DS1,DS2}, which in turn
reduces to the difference of the superpotentials,
\beq
\varepsilon_{ij} = 2[{\cal W}({\cal M}_j) -
{\cal W}({\cal M}_i)]\, ,
\label{sat}
\eeq
where the subscript ij marks the transition from the
$i$-th to $j$-th vacuum.
It is obvious then that if  a family
of the BPS domain walls 13 exist,
all walls from this family are degenerate;
their energy density is
\beq
{\cal E}(R) \equiv \varepsilon_{13} = \varepsilon_{12}+\varepsilon_{23}\, .
\label{sat2}
\eeq
The BPS domain walls 12 and 23 are the ``basic'' components
comprising all walls from the 13 family.
As a consequence of Eq. (\ref{sat}) they do not interact.

The fact of existence of a family of solutions in the 
13 transition can be easily established by inspecting the creek
equations \cite{CS} (see also \cite{Rey}) defining the BPS domain walls.
There is no need in finding the actual solutions of these
equations. The creek equations  can be interpreted
as complexified  equations of motion of a high viscosity fluid (whose inertia
can be neglected) on a profile associated with ${\cal W}$.
 For the real
superpotentials and real solutions  this interpretation is
especially simple, since complexification
becomes redundant, the profile is   given just by
 $-{\cal W}$ (with the above conventions regarding the ordering),
and a very rich physical intuition everyone has in 
this type of motion allows one to immediately see whether or not
a family of solutions exists in the given transition,
by a simple examination of the profile of 
 $-{\cal W}$. 

Although the above assertions are general,
I will illustrate them in the  generalized WZ models
describing dynamics of two chiral superfields.
There is no doubt, however, that under the same circumstances
degenerate families of the domain walls, with hidden parameters
and a variety of zero modes corresponding to
a change in the wall internal structure, appear in any model,
including strongly coupled gauge models. This may have important implications
for the domain walls in supersymmetric Yang-Mills theories \cite{KSS}.
I will not dwell on this topic in the present paper,
leaving it for future  publications.

The minimal WZ model, with one chiral superfield
and renormalizable superpotential,  generically has only one pair
of vacua, and a single  BPS wall, with no hidden parameters
\cite{DS1,CS}. The only bosonic zero mode existing in this
model is that associated with the translation of the wall as a whole.
Thus, this model is uninteresting
for our current purposes.
To reveal the phenomenon it is necessary to consider
models with two or more chiral superfields.
Since all essential ingredients appear already
at the level of two-field models, we will
limit ourselves to two chiral superfields.

\section{General Observations}

To explain the essence of the problem 
it is convenient to start from a non-supersymmetric system 
of two real scalar fields, $\phi$ and $\chi$,
\beq
{\cal L} = \frac{1}{2}\left[ 
(\partial_\mu\phi )^2 + (\partial_\mu\chi )^2
-V(\phi ,\chi )
\right]\, .
\eeq
For a short while we will forget about the
$\phi\chi$ coupling and consider decoupled fields.
We will incorporate a $\phi\chi$ interaction term
later. Assume that the self-interaction potential
has a double-well shape,
\beq
V_0 = \left(\frac{\mu^2}{\lambda} - \lambda\phi^2 \right)^2
+ \left(\frac{m^2}{g} - g\chi^2 \right)^2\, ,
\label{4}
\eeq
where $\mu$ and $m$ are the mass terms and $\lambda$ and $g$
are the coupling constants. This potential has four classical minima,
$\{\phi ,\chi \}_{*} $,
\beq
{\cal M}_1 =\left\{-\frac{\mu}{\lambda},
-\frac{m}{g} \right\},\,\,\,
{\cal M}_2 =
\left\{\frac{\mu}{\lambda},
-\frac{m}{g} \right\},\,\,\,
{\cal M}_3 =
\left\{-\frac{\mu}{\lambda},
\frac{m}{g} \right\},\,\,\,
{\cal M}_4 =
\left\{\frac{\mu}{\lambda},
\frac{m}{g} \right\}. 
\label{clamin}
\eeq
(It has also a maximum at the origin.) 
The field configuration interpolating between 
${\cal M}_1$ and ${\cal M}_2$
is the domain wall of the $\phi$ field,
\beq
\phi = \frac{\mu}{\lambda} \tanh \mu (z-z_0)\, ,
\,\,\, \chi = -\frac{m}{g}\, ,
\label{raz}
\eeq
while that interpolating between ${\cal M}_2$ and ${\cal M}_4$
is the domain wall of the $\chi$ field,
\beq
\phi = \frac{\mu}{\lambda}\, , \,\,\,   \chi = \frac{m}{g}
\tanh m (z-\zeta_0)\, ,
\label{dva}
\eeq
where $z_0$ and $\zeta_0$ are the centers of the 
corresponding walls. 

Finally, interpolating between the
first and the fourth minima is a superposition of two previous walls,
\beq
\phi = \frac{\mu}{\lambda} \tanh \mu (z-z_0)\, ,
\,\,\,  \chi = \frac{m}{g}
\tanh m (z-\zeta_0)\, .
\label{double}
\eeq

The double-wall configuration (\ref{double})
has two collective coordinates, $z_0$ and $\zeta_0$,
-- not surprisingly, of course, 
since the fields $\phi$ and $\chi$ are
decoupled so far, and the total energy density residing in the
configuration (\ref{double}) does not depend on the
relation between $z_0$ and $\zeta_0$, and is equal to
$$
{\cal E}(z_0 , \zeta_0 ) = \varepsilon_{12} + \varepsilon_{24}
\, ,
$$
\beq
 \varepsilon_{12} =\frac{4\mu^3}{3\lambda^2}\, ,\,\,\,
\varepsilon_{24} =\frac{4m^3}{3g^2}\, .
\eeq
In other words, the two components of the domain wall
configuration (\ref{double}) do not interact with each other.
They experience neither attraction nor repulsion.
One can say that the domain wall
configuration (\ref{double})  is infinitely (continuously)
degenerate. 

The profile of the potential energy $V_0$
is depicted on Fig. 1. The range
of variation of $\{\phi ,\chi\}$
corresponding to Fig. 1 is given on Fig. 2.
The wall solutions interpolating from 
$A$ to $B$ and from $B$ to $D$ are obviously the
``basic'' components of the wall solution interpolating from
$A$ to $D$. They are obviously
unique in the sense
that the $AB$ and $BD$  wall trajectories on
Fig. 1 are unique. That's not the case
for the $AD$ wall solution.
 The latter has a hidden parameter --
a relative position of the components'  centers, $R =\zeta_0- z_0 $.
If $R=0$ the $AD$ wall trajectory runs right on top of the hill on Fig.
1. If $R\neq 0$ it deviates either to the right from the
top or to the left. There exists a continuous family of
 trajectories, with one and the same
${\cal E}$. On  Fig. 2 the parameter $R$
is  reinterpreted as
an angle $\gamma$ determining the direction of motion
in the initial moment of time. 

\begin{figure}
  \epsfxsize=8cm
  \centerline{\epsfbox{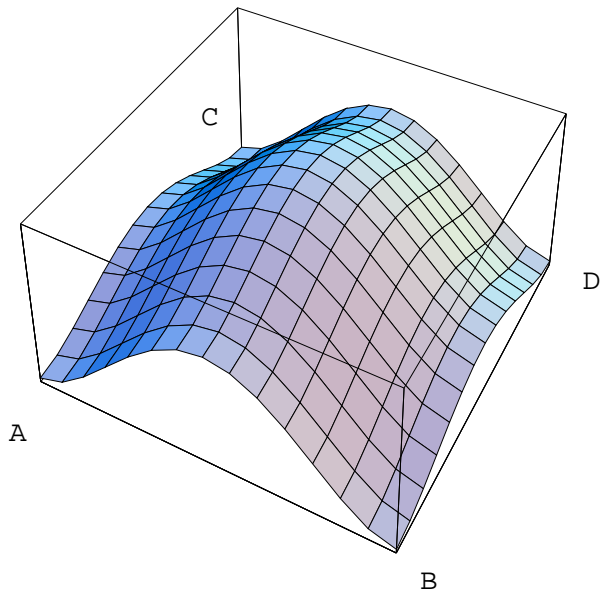}}
  \caption{The potential of the two-field model given in
Eq. (\ref{4}) for the following values of
the parameters: $\lambda = g = 1$, $\mu = 1$, $m = 1.2$. The points
$A,B,C,D$ mark four vacua of
the model. The four minima $A$ to $D$ correspond to
${\cal M}_1$ to ${\cal M}_4$, see Eq. (\ref{clamin}). }
  \label{potentialone}
\end{figure}   

\begin{figure}
  \epsfxsize=5cm
  \centerline{\epsfbox{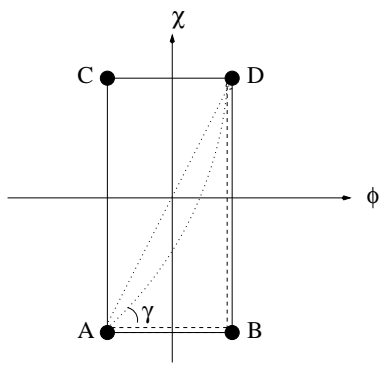}}
  \caption{The range of variation of the fields
$\phi$ and $\chi$ on the previous plot is shown by the solid line.
The four minima are depicted as closed circles. The dashed lines
show the wall trajectories $AB$ and $BD$, while
the dotted lines show two (out of infinitely many)
possible $AD$ trajectories.
 $\gamma$ is the injection angle
of the creek (at $z\rightarrow -\infty$).}
  \label{trajone}
\end{figure}

(I remind that the creek interpretation of the
equations defining the BPS wall \cite{CS,Rey}
implies that the variable $z$ is interpreted as
``time''. Correspondingly, differentiation over
$z$ will be denoted by dot, say $\dot\phi$ is the
$\phi$ component of the ``velocity'', $\dot\chi$
is the $\chi$ component, and so on. The angle $\gamma$
determines the direction of the ``velocity'' vector,
$\gamma = \mbox{arccot} (\dot{\phi}/\dot{\chi})$.
For instance, for the solution (\ref{double})
the angle of injection at the initial moment
of time, $z\rightarrow -\infty$, is
$\tan \gamma = \mu^2 m^{-2} g\lambda^{-1}\exp 2(\zeta_0 - z_0)$.) 

The degeneracy is {\em immediately} lifted, generally speaking,
once one switches on an interaction between
$\phi$ and $\chi$. Gone with this degeneracy is the
existence of the hidden parameter and a 
continuous family of the $AD$
trajectories.

Indeed, consider a typical interaction term, say,
\beq
\Delta V = \alpha \left(\chi^2 -\frac{m^2}{g^2} \right)
\left(\phi^2 -\frac{\mu^2}{\lambda^2} \right)\, ,
\eeq 
to be added to $V_0$. The interaction is chosen
in such a way that the positions of the minima of $V$
are not shifted (Fig. 2). This is done for technical reasons only,
to facilitate calculations. We could have easily dealt with
any other interaction term. To simplify things further
we will work in the limit $\alpha\ll 1$. This is
a technical assumption too,  inessential for the final
 conclusion.

In the first order in $\alpha$
the change in the wall tension is
$$
\Delta{\cal E} = 
\alpha\frac{\mu^2 m^2}{\lambda^2 g^2}I\, ,
$$
\beq
I =  
\int_{-\infty}^{\infty}
dz \frac{1}{\cosh^2(\mu  z)}\,\, 
\frac{1}{\cosh^2(m (z-R))}\, .
\eeq
If $\alpha < 0$ an attraction between the basic
wall components  arises; the $AB$ and $BD$ walls
collapse, and the only wall solution connecting the points
$A$ and $D$ that persists runs exactly on top of the hill.
On the other hand if $\alpha$ is positive, on the contrary,
the basic components experience repulsion,
and strictly speaking, there is no $AD$ wall at all.
It exists only as a limiting superposition of the $AB$ and $BD$ walls,
located
infinitely far from each other, $R\rightarrow\infty$.
In the first case the angle $\gamma$ on Fig. 2 is $\arctan (\lambda m /g\mu )$,
in the second case it is either zero or $\pi / 2$.
In any case the collective coordinate associated with
$R$ disappears.

Even if the interaction term 
$\Delta V$ is fine-tuned in such a way that
classically $\Delta {\cal E} =0$,
a non-vanishing $\Delta {\cal E}$ inevitably emerges at
the quantum level, as a result of loop corrections,
ruining the degeneracy of the $AD$ 
 trajectories inherent to the decoupled fields.
There is no symmetry which would force
$\Delta {\cal E}$ to stay at zero in the non-supersymmetric case once
$\Delta V \neq 0$,
and it does not.

In contrast, it will be shown that
supersymmetric BPS walls are generically continuously degenerate.
In the models with two chiral superfields,
besides the overall wall center, there exists
one extra collective coordinate
even in the presence of the $\phi\chi$  coupling. It
 characterizes  the wall internal
structure, and is analogous to $R$
or the angle $\gamma$.

\vspace{0.3cm}

Passing to the discussion of the 
continuous degeneracy of the domain walls in the generalized WZ
models,
as in the non-supersymmetric example above,  it is instructive
to start  from two decoupled superfields.
The superpotential has the form
\beq
{\cal W}_0 (\Phi , X)
= \left( \frac{\mu^2}{\lambda}\Phi -\frac{\lambda}{3}\Phi^3\right )+
 \left( \frac{m^2}{g}X -\frac{g}{3}X^3\right )\, .
\label{spone}
\eeq
(I hasten to add that a $\Phi X$ coupling
will be introduced shortly.) If the lowest components of the superfields $\Phi$
and $X$ are denoted by $\phi$ and $\chi$, the extrema
of the superpotential (\ref{spone})
(i.e. the solutions of the equations
$\partial{\cal W}_0 /\partial\Phi =0$ and $\partial{\cal W}_0 /\partial X =0$)
 are the same as in Eq. (\ref{clamin}). The values of the superpotential
at the extrema are
$$
({\cal W}_0)_{*} = \mp \frac{2\mu^3}{3\lambda^2}\mp \frac{2m^3}{3g^2}\, .
$$
The profile of the function $-{\cal W}_0(\Phi , X)$ is shown
on Fig. 3. The first extremum, ${\cal M}_1$, is the maximum of this function,
${\cal M}_4$ is the minimum, ${\cal M}_{2,3}$ are the saddle points.

\begin{figure}
  \epsfxsize=8cm
  \centerline{\epsfbox{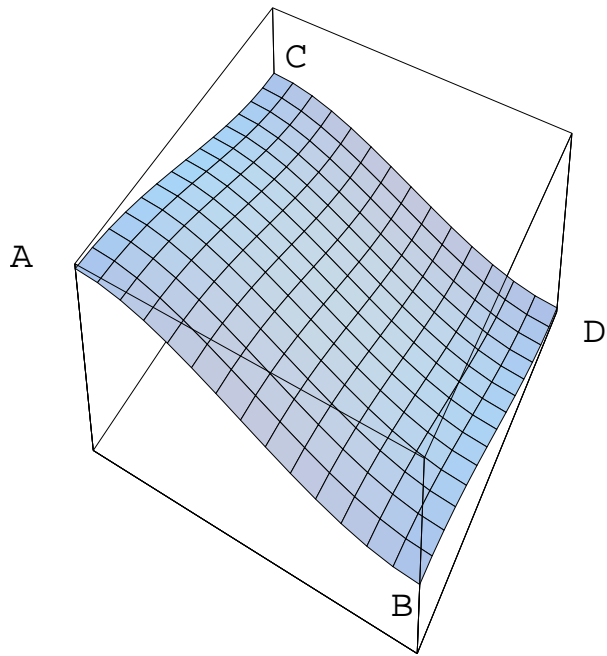}}
  \caption{The profile of the superpotential 
$-{\cal W}(\Phi , X)$, Eqs. (\ref{spone}), (\ref{typ}). The notations are
the same as on Figs. 1,2. }
  \label{supone}
\end{figure}   

The $AB$ and $BD$  walls exist, they are given by Eqs.
(\ref{raz}), (\ref{dva}). The corresponding
trajectories are unique.  The $AD$ domain wall (\ref{double})
represents a continuous family of trajectories, with an extra collective
coordinate, $R$ or $\gamma$,
$$
\gamma = \mbox{arccot} \left( \dot{\phi}/\dot{\chi} \right)_{z\rightarrow
-\infty} =
\mbox{arccot}\left[ \frac{\mu^2}{m^2}\, \frac{g}{\lambda}
(\cosh R + \sinh R)^2\right] \, .
$$
(Here $R=\zeta_0 - z_0$.)
 It can be viewed as a bound state of the
$AB$ and $BD$ walls, with the vanishing binding energy. 

So far, everything is in one-to-one correspondence with
the situation in the decoupled non-supersymmetric example.
Now comes a drastic distinction.

 Let us switch an interaction term
and show that: 

\vspace{0.1cm}
(i) a family of the $AD$ walls persists.
This family is degenerate since for any wall from the family
 ${\cal E} =
\varepsilon_{12}+ \varepsilon_{24}$, where
$$
\varepsilon_{12} = 2[{\cal W}({\cal M}_2) -
{\cal W}({\cal M}_1)]\, ,
$$
$$
\varepsilon_{24} = 2[{\cal W}({\cal M}_4) -
{\cal W}({\cal M}_2)]\, ;
$$

(ii) any interaction term
coupling $\Phi$ and $X$, 
which does not cause a ``catastrophic'' restructuring
of the profile ${\cal W}$, 
does guarantee the point (i). (I will
explain later what is meant by catastrophic.)

\vspace{0.1cm}

 As a matter of fact,
the equality ${\cal E} =
\varepsilon_{12}+ \varepsilon_{24}$ is a trivial consequence 
of the relation between the BPS wall tension and the central charge
in the transition at hand, similar to Eqs. (\ref{sat}), (\ref{sat2}).
We need to prove only that a continuous
family of the BPS trajectories, connecting the points
$A$ and $D$ (the extrema ${\cal M}_1$ and ${\cal M}_4$) exists.
In the absence of  coupling between  $\Phi$ and $X$,
the proof is explicit, see Eq. (\ref{double}).
When the interaction is switched on, the analytic form of the
 solution is unknown, but the fact of its existence
follows from 
the creek equations \cite{CS,Rey}
\beq
\dot{\vec\Phi} = \vec\nabla\overline{\cal W}\, .
\label{creek} 
\eeq
(Here $\vec\Phi$ is a generic set of the superfields;
under the rules of the game we have accepted,
 one can drop the bar over ${\cal W}$ on the right-hand side.)
We will prove two straightforward consequences of Eq.
(\ref{creek}).

\vspace{0.1cm}

I). An infinite number of the wall trajectories originate
from every maximum of $ (-{\cal W})$,
and infinitely many trajectories end up in every minimum.

II). Only one trajectory departs from every saddle point of
$ (-{\cal W})$, and only one arrives.

\vspace{0.1cm}

Needless to say that, since we are speaking
of maxima, minima and saddle points, we continue dealing,
 as previously,
only with the real solutions of the creek equations (\ref{creek}),
assuming all parameters in the superpotential to be real.
In the complex plane all extrema are saddle points, of course.

To prove the assertions (I) and (II)
above consider the profile ${\cal W}(\Phi , X)$
near the extremum points. Near the extrema
$$
{\cal W} = {\cal W}_* + P_2 (\delta\Phi ,\delta X)
$$
where 
$$
\delta\Phi = \Phi - \Phi_*\, , \,\,\,  \delta X
=X - X_*\, ,
$$
and $P_2$ is a homogeneous polynomial of the second order.
By a real rotation of the fields $\delta\Phi , \delta X$,
$$
\{\delta\Phi ,  \delta X\}\rightarrow
\{\Delta_1 ,\Delta_2\}
$$
one can always diagonolize $P_2$. In terms of the
diagonal variables $\Delta_{1,2}$ 
$$
P_2 = \frac{1}{2}A\Delta_1^2 +  \frac{1}{2}B\Delta_2^2
\, ,
$$
where $A,B$ are some constants,
and the creek equations take the form
\beq
\dot{\Delta}_1 = A\Delta_1\, , \,\,\, 
\dot{\Delta}_2 = B\Delta_2\, .
\label{smallcr}
\eeq
Both constants, $A$ and $B$ are positive near the maximum
of $-{\cal W}$, negative near the minimum, and one positive
one negative near the saddle points.
The appropriate asymptotic behavior of the trajectory
is $\Delta_{1,2}\rightarrow 0$ at $z\rightarrow
-\infty$ for the outgoing trajectory, and at
$z\rightarrow
\infty$ for the incoming trajectory. The solutions
of Eqs. (\ref{smallcr})
with the appropriate asymptotics are
\beq
\Delta_1 = C_1{\rm e}^{Az}\, , \,\,\, 
\Delta_2 = C_2{\rm e}^{Bz}
\label{minimax}
\eeq
for the trajectories 
leaving a  maximum or arriving at a minimum
of $-{\cal W}$. Here $C_{1,2}$ are arbitrary constants,
whose ratio determines $\gamma$. At the same time
for the trajectories attached to the saddle points
we have
\beq
\Delta_1 = C_1{\rm e}^{Az}\, , \,\,\, 
\Delta_2 = 0
\label{spI}
\eeq
and
\beq
\Delta_1 = 0\, , \,\,\, 
\Delta_2 = C_2{\rm e}^{Bz}\, .
\label{spII}
\eeq
The first one leaves a saddle point, the second arrives 
(I assume for definiteness that $A>0$, $B<0$.)
It is quite obvious that in Eq. (\ref{minimax})
a continuous parameter emerges, while there is
no such freedom in the case of Eqs. (\ref{spI}),
(\ref{spII}).

\vspace{0.1cm}

Not every trajectory leaving a maximum will end up at a minimum
(or at a saddle point,  a special case),
thus generating a legitimate BPS wall.
Some trajectories will lead to abysses,
yielding no BPS-saturated domain
wall solutions. In other words, there
are global constraints on the angle $\gamma$.
These constraints become clear from a visual examination of the
profile of $-{\cal W}$. Thus, in the trivial case of Eq.
(\ref{spone}) the boundary values of the angle
are $\gamma_* = 0$ and $\gamma_* =\pi /2$.
Introducing an interaction
between $\Phi$ and $X$ we shift the vacuum values of the
fields $\Phi_* , X_*$, the corresponding 
values of the superpotential (determining the central charges),
the boundary values $\gamma_*$, but as long
as the interaction term does not cause
a ``geographical'' disaster, the continuous degeneracy
of the $AD$ wall family will survive, the model 
will  support a unique trajectory for the
$AB$ and $BD$ walls, and a continuous family for the
$AD$ walls. 

A typical interaction is  
\beq
\Delta {\cal W} = 2\alpha\Phi X
\, , \,\,\, {\cal W} = {\cal W}_0 + \Delta {\cal W}\, .
\label{typ}
\eeq

The coupling between $\Phi$ and $X$ distorts details of the profile,
as compared with the
decoupling  limit, but the gross features remain the same:
one maximum, one minimum, two saddle points.
The maximum of $-{\cal W}$ is the highest point,
the saddle points are somewhat 
below, and the minimum of  $-{\cal W}$ is the lowest  point.
  Starting from the maximum,
the creek descends to either of the saddle points, from either of
 the saddle points
it descends to the minimum. Finally, there is
a family of trajectories connecting the maximum and the minimum directly.
What particular trajectory is chosen, depends on the angle $\gamma$
of the stream injection at the initial moment of time
(i.e. $z\rightarrow - \infty$). If $\alpha \ll |\mu - m|$
the boundary value of $\gamma$, instead of zero, becomes
$\gamma_* = \alpha |\mu - m |^{-1} + O(\alpha^2)$.

Other  couplings between $\Phi$ and $X$, not necessarily
reducing to Eq. (\ref{typ}),  are possible too.
The general pattern will continue to hold
until the ineteraction between $\Phi$ and $X$  becomes so  strong 
 that the 
gross features of the ``slope'' under consideration  change -- 
e.g. a new ``mountain ridge'' emerges
preventing the descent to the minimum, or the minimum raises up to the level
 of the maximum, and so
on. This can only happen under special conditions, 
at $\alpha \sim |\mu - m |$. This catastrophic
restructuring  is a different story, however, which will  not be  touched 
in the present paper. As long as the coupling  between $\Phi$ and $X$
does not change the overall general pattern of the extrema on the ``slope'',
a continuous family of the $AD$ walls will exist.

\section{Elaborating a Specific Example.}

To get further insight on the impact the
continuosly degenerate BPS wall families may have,
it is instructive to work out  particular models. Therefore, I choose
a concrete coupling between $\Phi$ and $X$,
and rewrite the two-field model at hand in a slightly different form
by passing to new superfields (which I will still continue  calling
 $\Phi$ and $X$),
\beq
{\cal W} = \frac{m^2}{\lambda}\Phi -\frac{\lambda}{3}\Phi^3
-\alpha\Phi X^2\, .
\label{shifmod}
\eeq
The four extrema $\{\Phi , X\}_{*}$
are
\beq
{\cal M}_1 = \{ -\frac{m}{\lambda}, 0\}\, , \,\,\,
{\cal M}_{2,3} = \{ 0,  \pm\frac{m}{\sqrt{\lambda\alpha}}\}\, ,\,\,\,
{\cal M}_4 = \{ \frac{m}{\lambda}, 0\}\, .
\eeq
The values of the superpotential at extrema are
\beq
{\cal W}({\cal M}_1)=-\frac{2}{3}\frac{m^3}{\lambda^2}\, , \,\,\,
{\cal W}({\cal M}_{2,3})=0\, , \,\,\,
{\cal W}({\cal M}_4)=\frac{2}{3}\frac{m^3}{\lambda^2}\, .
\eeq
The profile of the corresponding function $-{\cal W}$
is shown on Fig. 4, while the scalar
potential in the model at hand
is presented on Fig. 5.  The essential points are explained on Fig. 6.
It is assumed that $\alpha < \lambda$. As we will see shortly,
the relation between $\alpha$ and  $ \lambda$ is important.

\begin{figure}
  \epsfxsize=8cm
  \centerline{\epsfbox{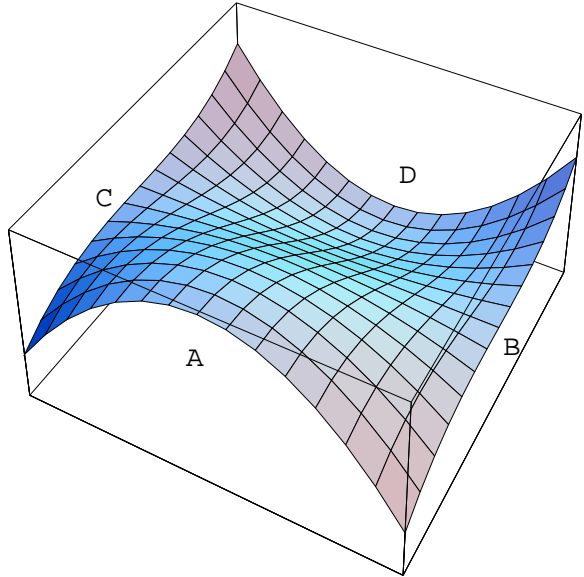}}
  \caption{The profile of the superpotential
$-{\cal W}$ in the model (\ref{shifmod})  for the following values of
the parameters: $\lambda = m = 1$, $\alpha = 0.49$. The points
$A,B,C,D$ mark four vacua of
the model: $A$ is maximum of $-{\cal W}$  corresponding to
${\cal M}_1$, $D$ is minimum corresponding to ${\cal M}_4$,
$B,C$ are saddle points ${\cal M}_{2,3}$. }
  \label{suptwo}
\end{figure}   

\begin{figure}
  \epsfxsize=7cm
  \centerline{\epsfbox{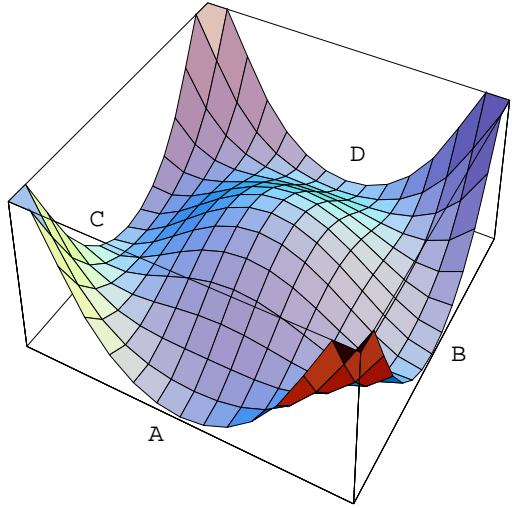}}
  \caption{The scalar potential in the same model.}
  \label{potentwo}
\end{figure}   

\begin{figure}
  \epsfxsize=6cm
  \centerline{\epsfbox{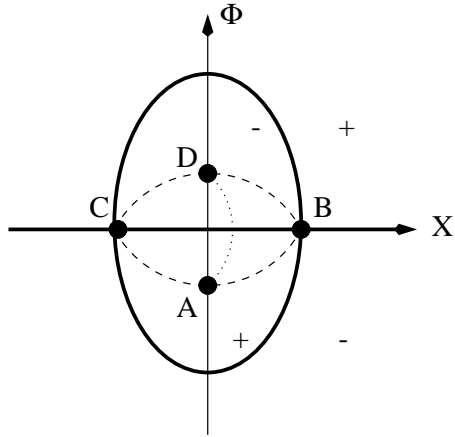}}
  \caption{The map of
$\Phi$ and $X$, for  the previous plots, with the level lines.
The thick solid lines denote zero of
the superpotential $-{\cal W}$, Eq. (\ref{shifmod}). The
regions of the positive height are marked by pluses, the regions of the negative height by minuses.
The dashed lines denote the trajectories of the
BPS walls coming to (or leaving from) the saddle points.
The dotted line is one of (infinitely many)  possible $AD$ walls.}
  \label{expl}
\end{figure}   

At $\alpha\neq\lambda$ the only apparent  symmetry of the model
(\ref{shifmod}) (additionally to supersymmetry)
 is a discrete $Z_4$,
\beq
\Phi \rightarrow \pm \Phi \,\,\, \mbox{and}\,\,\, X\rightarrow
\pm X \, .
\label{Z4}
\eeq
This symmetry connects the   vacua ${\cal M}_1$ and 
${\cal M}_4$, or   ${\cal M}_2$ and 
${\cal M}_3$:  ${\cal M}_1$ is physically
equivalent to ${\cal M}_4$ while ${\cal M}_2$ equivalent to 
${\cal M}_3$. $Z_4$
is sponatneously broken down to $Z_2$  in any of the four vacuum states.
 No symmetry relates  ${\cal M}_1$ to  ${\cal M}_2$.

As previously, $A$ marks the maximum, $D$ the minimum,
and $B,C$ mark the saddle points. The walls $AB$ and $AC$
are equivalent, and so are the walls $BD$ and $CD$.
The domain walls $AD$ and $BC$ are
different. The first one is BPS,
while the second is non-BPS;
their tensions do not coincide.

 The ellipse 
depicted on Fig. 6 by a thick solid line, as well as the  horizontal axis,
also depicted by  a thick solid line, are level lines --
they give the zero of the superpotential. Pluses and minuses
indicate the height of $-{\cal W}$
in the corresponding regions (positive or negative). The dashed line
$BACD$ is the boundary of the region
where a continuous family of the degenerate $AD$ trajectories
lies. Any trajectory leaving the point $A$ with the ``velocity''
directed in the lower half-plane will end up in abyss, while those
with the ``velocity'' in the upper half-plane will arrive at 
 the point
$D$. One of such trajectories is depicted by
the dotted line. The corresponding wall tension is
\beq
{\cal E} 
=\frac{8}{3}\frac{m^3}{\lambda^2}\, .
\eeq
 Two trajectories are exceptional;
they lead from $A$ to $C$ or $B$. The energy density of these walls
is
\beq
{\varepsilon} 
=\frac{4}{3}\frac{m^3}{\lambda^2}\, .
\label{bpset}
\end{equation}

The dashed line $BAC$ is the edge of the mountain ridge,
while the dashed line $CDB$ is the bottom of a valley.
By inspecting the matrix of the second derivatives
of ${\cal W}$ one readily convinces oneself
that the dashed line is horizontal at  the points
$A$ and $D$, while it approaches  the saddle points
$B$ and $C$  at the angles $\pm \pi / 4$. 
It is pretty obvious that the creek leaving $B$ at
$-\pi /4$ will arrive at $D$. If $\alpha <
\lambda /2$, the boundary trajectories from the $AD$
family can be found analytically,
\beq
\Phi = \frac{m}{\lambda}\tanh (Mz)\, ,\,\,\,
X = \pm \frac{m}{\sqrt{\alpha\lambda}}
\sqrt{1-\frac{2\alpha}{\lambda}}\,\, \frac{1}{\cosh (Mz)}\, ,
\label{predtr}
\eeq 
where $$
M = \frac{2\alpha m}{\lambda}.
$$
 For these trajectories
at $z\rightarrow -\infty$ the ``velocities'' are horizontal.

Instead of analyzing the creek equations, one could
prove the existence of the continuosly degenerate family of the $AD$ walls
in an indirect way, by counting the fermion zero modes
using the index theorem \cite{index}. A symmetric solution of the creek
equation,
\beq
X= 0\, , \,\,\, \Phi = \frac{m}{\lambda }\tanh (mz)
\label{SS}
\eeq
obviously exists. Now, if one calculates
the matrix of the second derivatives (the fermion mass
matrix) $\partial^2 {\cal W}/\partial\Phi_i\partial\Phi_j$
on the solution (\ref{SS}), this matrix is 
diagonal,
$$
\partial^2 {\cal W}/\partial\Phi_i\partial\Phi_j = -2\mbox{diag}
\{\lambda\Phi , \alpha\Phi\}\, ,
$$
with {\em both} eigenvalues changing sign along the trajectory
(\ref{SS}). From the index theorem \cite{index}
we then learn of the existence of two fermion zero modes.
Since the solution (\ref{SS}) preserves 1/2 of supersymmetry,
each fermion zero mode must have a boson partner.
Thus, we must have two boson zero modes. One is associated with a shift
of the wall center, another reflects the possibility of shifting
the trajectory along the ``slope'' (i.e. changing the
internal structure of the wall) without changing the tension.

\subsection{Non-BPS wall connecting the saddle points
${\cal M}_2$ and ${\cal M}_3$}

Since the points $B$ and $C$ both lie
at zero of the superpotential, there is
no BPS wall connecting them \cite{DS1,CS}.
A non-BPS wall  exists.
The corresponding value of $\Phi =0$,
while $X(z)$ satisfies the second-order
equation
\beq
\frac{d^2 X}{dz^2} = -2\alpha X \left( 
\frac{m^2}{\lambda} - \alpha X^2 \right).
\label{soeq}
\eeq
Its solution is
\beq
X = \frac{m}{\sqrt{\lambda\alpha}}\tanh (Mz)\, , \,\,\, M =
\sqrt{\frac{\alpha}{\lambda}}m\, .
\label{soeq2}
\eeq
A straightforward calculation of the tension of the $BC$ wall yields
\beq
\tilde{\cal E}= \sqrt{\frac{\lambda}{\alpha}}\,\, \frac{8}{3}\,
\frac{m^3}{\lambda^2} =  \sqrt{\frac{\lambda}{\alpha}}\,\,
2\varepsilon \, .
\label{edbcw}
\eeq
At $\alpha < \lambda$ the energy density of the
non-BPS wall (\ref{soeq2}) is higher than the sum of the energy densities
of the BPS walls connecting $BD$ and $DC$, see Eq. (\ref{bpset}).
The wall (\ref{soeq2}) is classically unstable
with respect to the decay into two BPS walls $BD$ and $DC$,
separated by an infinite $\Delta z$  interval. How the 
instability begins to develop is clearly seen
from Fig. 5. If we start from the solution (\ref{soeq2}),
with $\Phi =0$, it is energetically expedient to
push the trajectory away from the top of the hill
in the $\Phi$ direction. Quantitatively,
one can analyse the Hamiltonian for $\Phi$ in the background
 (\ref{soeq2}), assuming that $\Phi (z)$
is small, i.e. keeping only the quadratic terms
in $\Phi$ and omitting higher orders. The mode equation for 
$\Phi$ takes the form
\beq
\left\{
-\frac{d^2}{dz^2} +M^2 \left[ 4 - \left( 4+\frac{2\lambda}{\alpha}\right)
\, \frac{1}{\cosh^2 (Mz)}
\right]
\right\}\Phi_n (z) = E_n \Phi_n (z)
\label{mode}
\eeq
with the boundary conditions
$$
\Phi_n (z\rightarrow \pm \infty ) = 0\, .
$$
The parameter $M$ is the same as in Eq.  (\ref{soeq2}).

At $\alpha < \lambda$ the lowest mode $\Phi_0$ is negative,
$E_0 <0$. This means that allowing the wall trajectory to slide down
in the direction of $\Phi_0$,
$$
\Phi \sim \Phi_0\, ,
$$
we make the energy density of the $BC$ wall lower than that
in Eq. (\ref{edbcw}). This is the way the instability
in Eq. (\ref{soeq2}) starts. The evolution of the instability ends
when the wall (\ref{soeq2}) breaks into two well-separated pieces,
two BPS walls connecting ${\cal M}_2$ to ${\cal M}_4$
and ${\cal M}_4$ to ${\cal M}_3$, respectively.

If $\alpha > \lambda$, on the contrary, the above two BPS walls
are attracted to each other. They form a stable bound state,
a non-BPS wall   (\ref{soeq2}), connecting ${\cal M}_2$ to ${\cal M}_3$
directly. The wall tension $\tilde{\cal E}$ is smaller
than the sum of the tensions of the $BD$ and $DC$ walls.

Note, that the  tensions of the BPS walls are calculated exactly,
while those of the non-BPS walls, generally speaking,
receive corrections due to quantum loops. If the coupling constants
are small, these corrections are small too, and can be neglected
everywhere except in the immediate vicinity of the point
 $\alpha = \lambda $. 

The point $\alpha = \lambda $ is special. At this point the tension of the
non-BPS wall $BC$ is exactly equal to the sum
of the tensions of the $BD$ and $DC$ walls and equal
to the tension of the BPS wall $AD$,
\beq
\tilde{\cal E} = 2\varepsilon = {\cal E}\, .
\label{surpr}
\eeq
This is due to the fact that at $\alpha = \lambda $ the model
(\ref{shifmod}) degenerates into a system of two decoupled superfilds
$(\Phi \pm X)/\sqrt{2}$, and the $BC$ wall becomes
physically  identical to the $AD$ one.
Thus, although an additional symmetry emerges at
$\alpha = \lambda $, this limit is uninteresting.

\subsection{Integrating out a heavy field}

In many applications one has to deal with effective
Lagrangians which are written  for light degrees of freedom
after one integrates out heavy degrees of freedom.
An example which is widely discussed now
is the effective Lagrangian for the supersymmetric
Yang-Mills theory \cite{VY,KS}. Here we show that,
integrating out  heavy fields, typically  one  
 erases any trace of the continuous degeneracy of the
BPS walls existing before the heavy degrees of freedom are eliminated.

Let us turn again  to the model (\ref{shifmod}), 
and consider the limit $\alpha \gg \lambda$. Then in the vacua
${\cal M}_1$ and ${\cal M}_4$ the field $X$ is much heavier
than $\Phi$,
\beq
\frac{M_X}{M_\Phi} = \frac{\alpha}{\lambda}\, .
\label{heavy}
\eeq
As a matter of fact, this ratio holds (almost) everywhere
along the trajectory connecting ${\cal M}_1$ and ${\cal M}_4$.
The only exception is at $\Phi = 0$.
Therefore, following a standard routine, one is tempted to integrate
out the field $X$ in order to obtain an effective Lagrangian
for the ``light'' field $\Phi$. The standard routine is based
on the Born-Oppenheimer procedure: one freezes the value
of $\Phi$, and for every given value finds an optimal value of $X$
minimizing the energy of the field configuration at hand.
In this way one finds that for all values of $\Phi$
(except $\Phi = 0$, but we will forget about this 
one ``singular'' point,
as it is commonly done) the corresponding optimal value of $X$ vanishes,
as a consequence of  the equation $\partial {\cal W} /\partial X = 0$.
Substituting this solution back to ${\cal W} (\Phi , X)$ given in Eq.
(\ref{shifmod}), we arrive at the effective Lagrangian
for the $\Phi$ field, representing nothing but the minimal WZ model.
As is well-known \cite{DS1,CS}, 
the wall solution in this model is unique. Thus, integrating out $X$
a l\`{a} Born-Oppenheimer we loose any possibility of
exploring the  continuous family of the BPS walls, which exists
in the full theory.
It is highly probable that a similar situation may take place in the
Veneziano-Yankielowicz effective Lagrangian \cite{VY}
(see also \cite{KSS}),
where an uncontrollable number of ``heavy'' degrees of freedom
is eliminated. Whether this is the case, and if so, what is
the dimension of the parameter space of the BPS walls
in the supersymmetric Yang-Mills theories remains an open question.

\vspace{0.3cm}

\subsection*{Acknowledgments} 

I
would like to thank M. Voloshin
for a discussion.
This work was done during my stay at Institut f\"{u}r
 Theoretische Physik III, Universit\"{a}t
Erlangen-N\"{u}rnberg. I am grateful to F. Lenz
and other members of the group for kind hospitality.

This work was supported in part by DOE under the grant number
DE-FG02-94ER40823 and by Alexander von Humboldt-Stiftung.


\newpage

\begin{thebibliography}{99}

\bibitem{DS1}
G. Dvali and M. Shifman, hep-th/9611213 [Nucl. Phys. B, to appear].

\bibitem{DS2}
G. Dvali and M. Shifman, {\it Phys. Lett.} {\bf B396} (1997) 64.

\bibitem{KSS}
A. Kovner, M. Shifman and A. Smilga, hep-th/9706089 [Phys. Rev. D,
submited].

\bibitem{CS}
B. Chibisov and M. Shifman, hep-th/9706141 [Phys. Rev. D,
submited].

\bibitem{VY}
G. Veneziano and S. Yankielowicz,  {\it Phys. Lett.} {\bf B113} (1982) 231.

\bibitem{KS}
A. Kovner and M. Shifman,  hep-th/9702174 [Phys. Rev. D,
to appear].

\bibitem{WZ}
J. Wess and B. Zumino,  {\it Phys. Lett.} {\bf B49} (1974) 52.

\bibitem{Rey}
E.R.C. Abraham and P.K. Townsend, {\it Nucl. Phys.} {\bf B351} (1991) 313;\\
M. Cveti\v{c}, F. Quevedo and S.-J.  Rey,
{\it Phys. Rev. Lett.} {\bf 67} (1991) 1836.

\bibitem{index}
R. Jackiw and C. Rebbi, {\it Phys. Rev.} {\bf D13} (1976) 3398;\\
E. Weinberg, {\it Phys. Rev.} {\bf D24} (1981) 2669.


\end{thebibliography}
\end{document}